\documentclass[sn-basic]{sn-jnl}

\usepackage{graphicx}
\usepackage{multirow}
\usepackage{amsmath,amssymb,amsfonts}
\usepackage{amsthm}
\usepackage{mathrsfs}
\usepackage[title]{appendix}
\usepackage{xcolor}
\usepackage{textcomp}
\usepackage{manyfoot}
\usepackage{booktabs}
\usepackage{algorithm}
\usepackage{algorithmicx}
\usepackage{algpseudocode}
\usepackage{listings}
\usepackage{natbib}
\usepackage{multicol}

\begin{document}

\title[Strategies and statistical evaluation of Italy's regional model for COVID-19 restrictions]{Strategies and statistical evaluation of Italy's regional model for COVID-19 restrictions}

\author[1]{\fnm{Giuseppe} \sur{Drago}}\email{giuseppe.drago01@unipa.it}
\author*[1]{\fnm{Giulia} \sur{Marcon}}\email{giulia.marcon@unipa.it}
\author[1]{\fnm{Alberto} \sur{Lombardo}}\email{alberto.lombardo@unipa.it}
\author[1]{\fnm{Giuseppe} \sur{Aiello}}\email{giuseppe.aiello03@unipa.it}

\affil[1]{\orgdiv{Department of Engineering}, \orgname{University of Palermo}, \orgaddress{\street{Viale delle Scienze}, \city{Palermo}, \postcode{90128}, \country{Italy}}}

\abstract{This study presents a comprehensive assessment of the Italian risk model used during the COVID-19 pandemic to guide regional mobility restrictions through a colour-coded classification system. The research focuses on evaluating the variables selected by the Italian Ministry of Health for this purpose and their effectiveness in supporting public health decision-making. The analysis adopts a statistical framework which combines data reduction and regression modelling techniques to enhance interpretability and predictive accuracy. Dimensionality reduction is applied to address multicollinearity and simplify complex variable structures, while an ordinal regression model is employed to investigate the relationship between the reduced set of variables and the colour regional classifications. Model performance is evaluated using classification error metrics, providing insights into the adequacy of the selected variables in explaining the decision-making process. 
	Results reveal significant redundancy within the variables chosen by the Italian Ministry of Health, suggesting that excessive predictors may compromise information. To address this, the study proposes refined and robust predictive models for regional classification, 
	offering a reliable tool of the proposed framework and to support public health decision-makers.
	This study contributes to the ongoing development of quantitative methodologies aimed at improving the effectiveness of statistical models in guiding public health policies. The findings offer valuable insights for refining data-driven decision-making processes during health crises and improving the quality of information available to policymakers.}

\keywords{Risk Model, Data Quality Management, COVID-19, Dimensionality Reduction}

\maketitle

\section{Introduction}
\label{introduction}
The recent SARS-CoV-2 family of viruses known through COVID-19 which has become a global disease with devastating effects on the world economy and lifestyles. In this context, states around the world have attempted to counteract the effects of the pandemic through restrictive measures such as lockdown, which on  one hand have reduced its spread in terms of contagion but, on the other hand, have devastated the economy. In fact, although lockdown allows the contagion to be contained, it also has serious repercussions on the economic and financial activity of the country that makes use of it. For this reason, companies, governments, central banks and financial market regulators have been called upon to make major decision-making choices in order to adopt measures to cope with the impact of the crisis on economic activity and the financial system. Such measures require appropriate decision-making and allocation processes based on robust models; otherwise, one runs the risk of basing the decision-making process on non-rational choices. In fact, ``data inconsistency'' models can lead to ineffective decisions, both in terms of protecting human health and in economic terms. In this regard, it is important to prevent the risks associated with the irrelevance and inaccuracy of the data contained in the information systems used for forecasting. According to \cite{1_bakhrushin2020risks}, experience in forecasting the development of the COVID-19 pandemic shows that primary data are not always suitable for direct application in mathematical models. During the COVID-19 pandemic in Italy, the country has used a color-coded system to assign different levels of restrictions and measures to different regions. The color-coding system used to assign a colour to each region is based on the level of risk posed by the spread of COVID-19, thus in terms of the severity of the situation. The color-coding system had been updated weekly and considered a range of factors, including the number of new cases, the number of people hospitalized with COVID-19, the number of deaths, and the overall healthcare system capacity. The system categorizes regions into three different zones with each zone assigned a different colour:
\begin{itemize}
	\item Red Zone (High Risk): The highest level of restrictions, typically used in areas with a high number of COVID-19 cases and hospitalizations; regions assigned the red color are considered to have a high risk of COVID-19 transmission and are subject to the most restrictive measures.
	\item Yellow Zone (Medium Risk): Lower level of restrictions, used in areas with a low number of cases and hospitalizations; regions assigned the yellow color are considered to have a moderate risk of COVID-19 transmission and are subject to intermediate restrictions.
	\item Green Zone (Low Risk): The lowest level of restrictions, used in areas with a very low number of cases and hospitalizations; regions assigned the green color are considered to have a low risk of COVID-19 transmission and are subject to relatively relaxed restrictions.
\end{itemize}
The exact restrictions associated to each colour can vary depending on specific circumstances in each region, but in general, regions assigned a higher risk level are subjected to more restrictive measures, such as limits on meetings, restrictions on travel and mandatory use of masks. The italian colour system was intented to provide a clear and concise way to communicate the level of risk posed by COVID-19 in each region and to guide decision-making about restrictions and pubblic health measures. However, the inventor of the ``colour theory'' against COVID-19 argues that there are too many parameters and too little transparency in the choices in the italian colour coding system \citep{2_website_repubblica}. 
The choice of the number of input variables is crucial within a model, as there is a risk of leaving a margin in the calculation formula by forcing technicians and politicians to make a random decision. Since the lockdown measures in Italy are decided through this model, this work aims to analyse and evaluate the relationships between the variables used by the Italian Ministry of Health to assign the colours to the single regions, used for the mobility restrictions during the COVID-19 pandemic period.  Specifically, the research questions are the following:
\begin{enumerate}
	\item Is it possible in the Italian colour coding system to reduce the dimensionality of such datasets, increasing interpretability and at the same time minimising information loss?
	\item Is the Italian colour coding system a consistent model?
\end{enumerate}
Section \ref{literature_analysis} summaries the most recent literature about COVID-19 model analysis.
The statistical methodology and data description are presented in section \ref{statistical_methodology}. Results are collected in section \ref{results} while last section provides a brief conclusion and discussion highlighting limitations and insights for feature research.
\section{Literature Analysis}
\label{literature_analysis}
Access to accurate outbreak prediction models is essential to obtain information on the likely spread and consequences of infectious diseases. Governments and other legislative bodies rely on the insights of prediction models to suggest new policies and to evaluate the effectiveness of implemented policies \citep{3_remuzzi2020covid}.There are several studies using Machine Learning (ML) methods used for outbreak prediction. In \cite{4_liang2020prediction} the authors used data from African Swine Fever A outbreaks and meteorological data from the WorldClim database choose the CfsSubset Evaluator-Best First feature selection method combined with random forest algorithms to build a prediction model. The accuracy of support vector machine, artificial neural-network and random-forest time series models in modelling influenza-like illness (ILI) and outbreak detection was investigated by \cite{5_tapak2019comparative}. After conducting analysis and forecasting to estimate Aedes populations, \cite{6_raja2019artificial} used the results to infer the possibility of dengue outbreaks at predetermined locations around the Klang Valley, Malaysia. In particular, the Bayesian Network machine learning technique was used. An artificial intelligence-based model, called the ANN-2Day model, is presented to predict, manage and ultimately eliminate the increasing risk of oyster norovirus outbreaks \citep{7_chenar2018development}. It was found that oyster norovirus outbreaks can be predicted with a lead time of two days using the ANN-2Day model and daily data from some environmental predictors. The 2-day lead time allows public health to plan management interventions.
These ML methods are limited to the basic methods of random forest, neural networks, Bayesian networks, naive Bayes, genetic programming, and classification and regression tree (CART). Although ML has long been established as a standard tool for modelling natural disasters and weather forecasting  \citep{8_choubin2019earth, 9_choubin2020spatial}, its application in modelling outbreaks is still in the early stages. This condition is even harsher in the case of COVID-19. Indeed, the latter has shown a non-linear and complex nature  \citep{10_ivanov2020predicting}. Furthermore, the outbreak shows differences with other recent epidemics, which calls into question the ability of standard models to provide accurate results  \citep{11_koolhof2020forecasting}. In addition to the many known and unknown variables involved in the spread, the complexity of population-level behaviour in various geopolitical areas and differences in containment strategies increased model uncertainty  \citep{12_darwish2020comparative}. Indeed, the observation of COVID-19 behaviour in different countries demonstrates a high degree of uncertainty and complexity \citep{13_zhong2020early}. 
Therefore, for epidemiological models to provide reliable results, they must be adapted to the local situation with a view to susceptibility to infection  \citep{14_layne2020new}. The challenge is to build models that can overcome the limitations of conventional models. Indeed, advancing accurate models with high generalisation capacity to model both regional and global pandemics is essential for decision making  \citep{15_reis2019superensemble}. Consequently, to overcome these challenges, many new models have emerged that introduce different assumptions to modelling. Several studies show the spread of COVID-19 and its effects on detection, diagnosis and transmission rates. Many mathematical models have been derived to predict the spread of COVID-19 and provided diagnostics. A first Monte Carlo simulation model was applied to show the transmission rate of COVID-19 with the help of an average daily reproduction number ($R_t$), considering several parameters, such as the number of daily cases and the number of confirmed versus dead cases \citep{16_kucharski2020early}. A recent study was conducted on COVID-19 diagnostics, using radiological images. Systematic reviews were collected from three different databases, PubMed, Scopus and Web of Science, providing information on which model provided the best accuracy in terms of sensitivity and specificity values \citep{17_ghaderzadeh2021deep}. Time series analysis predicts the future of a particular phenomenon using previous data. It is classified into three different classes \citep{18_khan2020toward}: statistical modelling, machine learning modelling and deep learning modelling for time series analysis. The proposed research concerns the development of an improved model for the prediction of COVID-19 cases with respect to time. It provides information on the diffusion pattern of COVID-19 by visualising its diffusion graph, helps policy makers to determine the best policies that are useful for improving the accuracy of forecasting and meaningful for controlling COVID-19. To predict the growth rate of COVID-19, researchers have used various approaches to achieve maximum accuracy \citep{19_khan2022artificial}. \cite{20_amar2020prediction} used an ML regression model to predict the rate of spread of COVID-19 in Egypt. They used seven different variants of the regression model and found the best approach with the lowest error value. The models predicted COVID-19 for the next 15 days with a dataset provided by the government. 
Several studies have been used to predict the future value of COVID-19.
\cite{21_shastri2020time} aim to predict the future conditions of new Coronaviruses to reduce their impact utilised databases \citep{22_covidindia, 23_covidusa}.
This is a comparative analysis based on deep learning of COVID-19 cases in India and the United States.
In \cite{24_mohammadi2020transformer} a self-attention-based transformer network with exceptional capability in time series forecasting is demonstrated. The goal is to forecast time series at large intervals.
In \cite{25_jin2021inter}, the author demonstrates that COVID-19 spreads at varying speeds and scales in distinct geographic regions, creating a new model that produces forecasts by analysing patterns on different time series data.
\cite{26_istaiteh2020} conducts a comparative analysis of four different models and promotes a worldwide forecasting tool that predicts confirmed cases of COVID-19 for the next seven days worldwide.
\cite{27_guptaa2021} demonstrates that the ``Prophet Forecasting Model'' is the best predictive method for predicting the active rate, mortality rate and cure rate compared to SVM and linear regression in the presence of large datasets characterised by a certain level of uncertainty. 
\cite{28_ardabili2020} predicted COVID-19 for Italy, China, Iran, Germany and the United States.  Cases of COVID-19 in the US using ML and statistical models are studied in \cite{29_cobb2020examining}
\section{Statistical Methodology}
\label{statistical_methodology}
The expected effectiveness of the Italian system of colour region assignment during the COVID-19 pandemic is a subjective matter that depends on the perspective and criteria used to evaluate it. Some may argue that the system was effective in managing the spread of the virus and preventing the healthcare system from being overwhelmed, while others may argue that the system was not stringent enough or was inconsistent in its application.
It is important to consider the trade-off between controlling the spread of the virus and maintaining economic activity and personal freedoms, as well as the availability of reliable data and the evolving understanding of the virus. It is also worth noting that the effectiveness of the colour region assignment system may have varied over time as the situation evolved and as the response was adjusted.
\par 	
However, the number of input variables used in the Italian model for colour region assignment during the COVID-19 pandemic was influenced by the need to make informed decisions about the spread of the virus and the response to it. The model was designed to provide a comprehensive picture of the situation, taking into account a wide range of variables that could impact the spread of the virus and the ability of the healthcare system to respond.
It is possible that the same information could have been obtained by using fewer input variables providing simpler and more manageable models without leading to a loss of information or less accurate predictions or decisions.
\par
In order to identify the most informative subset of variables, all combinations have been considered. For each combination, healthcare-related data have been used to run a principal component analysis (PCA) in order to select a lower number of components which explain almost the whole percentage of cumulative variability (at least 90\%). 
Every PC's set has been used as input set to estimate an ordinal regression model whose output variable is the region colour (three levels: `H' high, `M' medium, `L' low). Then, an explanatory analysis has been conducted on the misclassification error. 
\par
PCA has been used to reduce the dimensionality of the original variables and to remove the multicollinearity between them by transforming the variables into a set of uncorrelated component. Thus, the principal components capture the underlying structure of the predictors, while reducing the degree of linear association between them. In ordinal regression, the transformed principal components are used as the predictors in the model, instead of the original variables. This can help to stabilize the estimates of the regression coefficients and to improve the performance of the model. 
\par
Data processing and statistical analysis have been performed by using R software \cite{Rsoftware}. 

\subsection{Data Aggregation}
\par
The input variables used in the model include healthcare-related data such as the number of confirmed cases, the number of hospitalizations, the number of deaths, the availability of healthcare resources. Data have been collected from the \cite{dati} referring to the daily time interval from January 1$^{st}$, 2021 to December 31$^{st}$, 2021:
\begin{multicols}{2}
	\begin{itemize}
		\item[$X_1$:] Hospitalized with symptoms 
		\item[$X_2$:] ICU patients 
		\item[$X_3$:] ICU daily admissions 
		\item[$X_4$:] Home quarantine 
		\item[$X_5$:] Confirmed cases 
		\item[$X_6$:] Discharged healed 
		\item[$X_7$:] Deaths 
		\item[$X_8$:] Cases confirmed by PCR 
		\item[$X_9$:] Cases confirmed by RAT 
		\item[$X_{10}$:] Total cases 
		\item[$X_{11}$:] Increase in total cases (compared to the previous day) 
		\item[$X_{12}$:] People Tested	
		\item[$X_{13}$:] PCR 
		\item[$X_{14}$:] RAT 
		\item[$X_{15}$:] Total swabs 
		\item[$X_{16}$:] Increase in total swabs (compared to the previous day) 
	\end{itemize}	
\end{multicols}
The output variable $Y$, corresponding to the risk level expressed in terms of colours, has weekly frequency and has been collected from the same source and referring to the same time interval. \par
\begin{figure}[!h]
	\centering
	\includegraphics[width=\textwidth]{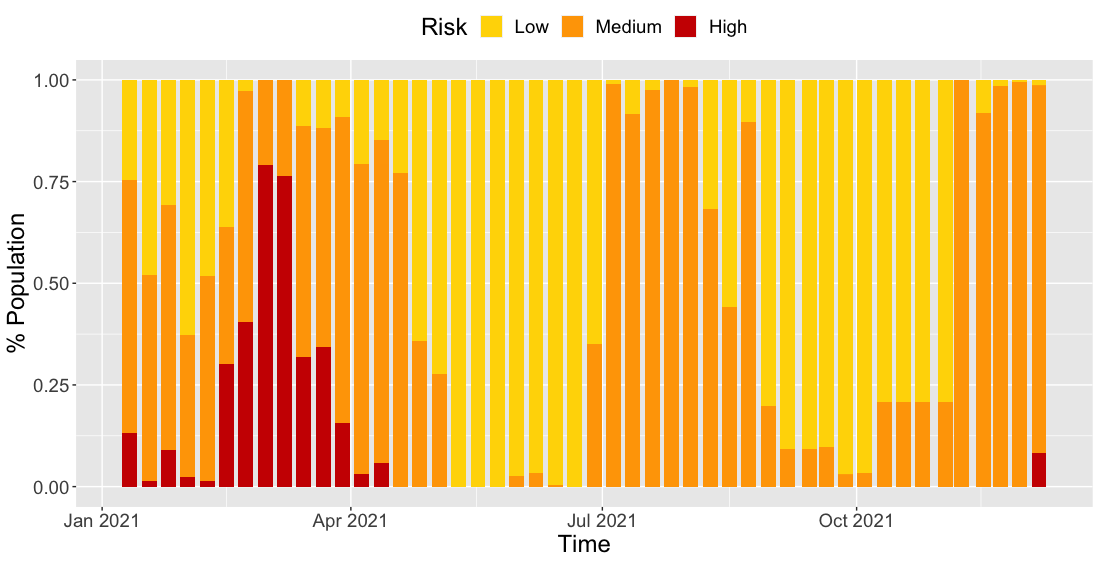}
	\caption{Risk levels in the Italian regions, expressed in terms of percentage of the Italian population characterized by the coloured markers yellow, orange and red, in the January 1$^{st}$, 2021 - December 31$^{st}$, 2021 period of time.}
	\label{fig:perc_color_pop}
\end{figure}

 
Intervals of time are measured in different units, then a weekly aggregation of the input time series data was conducted before performing both PCA and ordinal regression. Figure \ref{fig:perc_color_pop} shows the colour distribution over time through the percentage of Italian population corresponding to each coloured region.
%
\subsection{Principal Component Analysis}
\begin{figure}[!h]
	\centering
	\includegraphics[width=\textwidth]{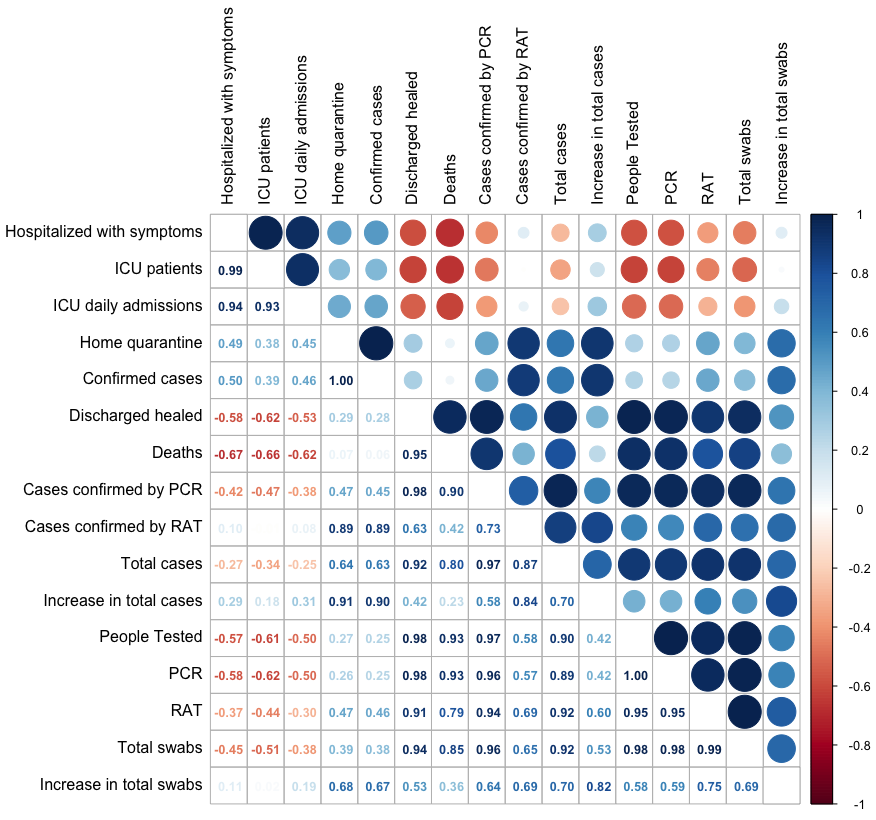}
	\caption{Correlation matrix of national data.}
	\label{fig:matrice_correlazione}
\end{figure}

Principal Component Analysis (PCA) is a technique that reduces the dimensionality of the independent variables by creating a new set of orthogonal variables, called principal components, that capture the most important variations in the data. PCA has been applied to the original scaled data in order to avoid multicollinearity in ordinal regression as input variables have quite strong linear correlations (see Fig. \ref{fig:matrice_correlazione}). 
In PCA, the trade-off between the cumulative percentage variance explained and the number of principal components is an important consideration. The cumulative percentage variance explained represents the proportion of the total variance in the data that is explained by the first few principal components. \par
In this context, PCA is used as a pre-processing step for ordinal regression, then the threshold of cumulative percentage variance explained has been fixed such that at least the 90\% is captured, for this reason the number of principal components may change. 
The inclusion of at most four principal components ensures that sufficient information is retained to accurately model the relationship between the predictors and the outcome variable.
%
\subsection{Ordinal Regression}
Ordinal regression is a statistical technique used to model the relationship between a dependent variable with ordered categories and independent variables. In order to address the independency of the covariates, a possible approach that has been used in this paper applies principal component analysis (PCA) to such covariates and then use the principal components as the input variables in an ordinal regression model.
\par
Prediction performances of the models estimated are expressed in terms of Misclassification Error which refers to the incorrect assignment of a dependent variable's value to an observation. The measure of the misclassification error is computed as the Error Rate, thus the proportion of observations that are incorrectly classified. 
\newline
For each region, all 65.535 combinations $C^n_k$ of the input variables (with $n = 1, \ldots, 16$ and $k = 1, \ldots, 16$) have been considered in order to find the best combination set of the input variables transformed into PC through the variables loadings $e_{ij}$ (see eq. \ref{eq: PCs}, $i, j, p = 1, \ldots, 16$) 
\begin{equation}
	\label{eq: PCs}
	\begin{array}{lll} 
		PC_1 & = & e_{11}X_1 + e_{12}X_2 + \dots + e_{1p}X_p \\ 
		PC_2 & = & e_{21}X_1 + e_{22}X_2 + \dots + e_{2p}X_p \\ 
		& & \vdots \\ 
		PC_p & = & e_{p1}X_1 + e_{p2}X_2 + \dots +e_{pp}X_p
	\end{array}
\end{equation}
which provides the minimum misclassification error in the ordinal regression models (eq. \ref{eq: or}, $r = 1, \ldots, p$, $\ell = 1, 2$). 
\begin{equation}
	\label{eq: or}
	\mbox{logit} \Big( P\left(Y \le \ell \right) \Big)  = \eta_\ell + \beta_1 PC_1 + \cdots + \beta_r PC_r \hspace{1cm}  
\end{equation}
\section{Results and discussion}
\label{results}
The explanatory analysis conducted through the PCA has been provided that, among all different combinations, the number of components which explain almost the whole percentage of cumulative variability (at least 90\%) is distributed in the following way: 1 PC has been selected 0.94\% of times, 2 PC 75.54\%, 3 PC 23.07\% and 4 PC 0.45\%. 
\par
Results have been synthesized through the misclassification error.  
\begin{figure}[!h]
	\centering
	\includegraphics[width=\textwidth]{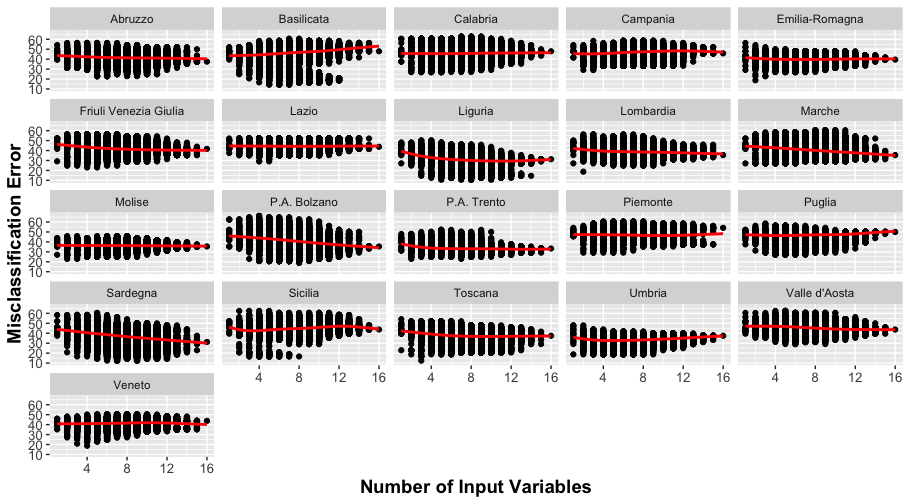}
	\caption{Misclassification errors and number of input variables.}
	\label{fig:scatter_misclerr_idi}
\end{figure}
Figure \ref{fig:scatter_misclerr_idi} highlights how the average trend of the misclassification error does not decrease when increasing the number of input variables. On the other hand the high variability around the average values, for each region, suggests the possibility of selecting best models in terms of lower misclassification errors and number of input variables. \\
Most of the regions, except for P.A. Bolzano and Sardegna, show lower misclassification errors corresponding to the ordinal regression models where the covariates are the principal components involving less than five input variables. 
This may demonstrate that the Italian system is not efficient. Such analysis suggests the possibility of reducing the dimensionality of input variables to increase its interpretability while minimizing the loss of information. Thus, assuming that the attribution of colours provided by the National system is correct, it would be possible to simplify each regional model by selecting only a reasonable subset of input variables. \par
\begin{figure}[!h]
	\centering
	\includegraphics[width=.7\textwidth]{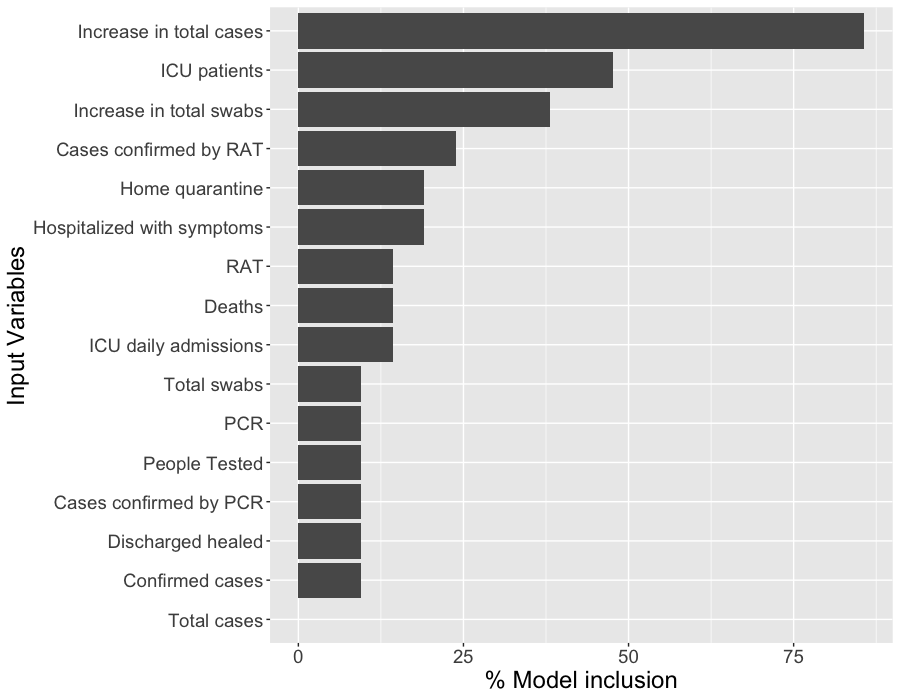}
	\caption{Percentage of model inclusion for each variable among the models with minimum misclassification errors.}
	\label{fig:vars}
\end{figure}

\begin{figure}[!h]
	\centering
	\includegraphics[width=0.3\textwidth, page=1]{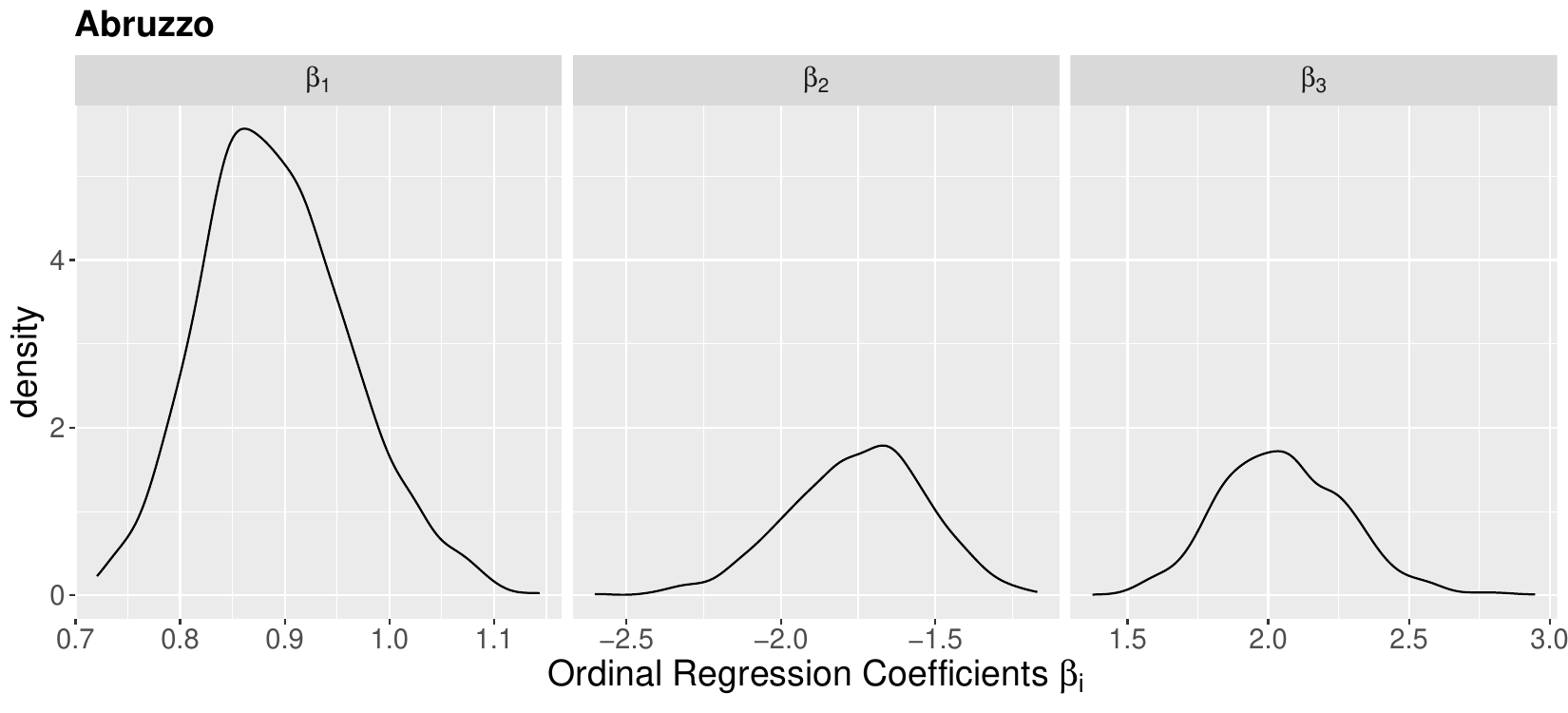}
	\includegraphics[width=0.3\textwidth, page=2]{fig5_jack_coef_free.pdf}
	\includegraphics[width=0.3\textwidth, page=3]{fig5_jack_coef_free.pdf}
	\includegraphics[width=0.3\textwidth, page=4]{fig5_jack_coef_free.pdf}
	\includegraphics[width=0.3\textwidth, page=5]{fig5_jack_coef_free.pdf}
	\includegraphics[width=0.3\textwidth, page=6]{fig5_jack_coef_free.pdf}
	\includegraphics[width=0.3\textwidth, page=7]{fig5_jack_coef_free.pdf}
	\includegraphics[width=0.3\textwidth, page=8]{fig5_jack_coef_free.pdf}
	\includegraphics[width=0.3\textwidth, page=9]{fig5_jack_coef_free.pdf}
	\includegraphics[width=0.3\textwidth, page=10]{fig5_jack_coef_free.pdf}
	\includegraphics[width=0.3\textwidth, page=11]{fig5_jack_coef_free.pdf}
	\includegraphics[width=0.3\textwidth, page=12]{fig5_jack_coef_free.pdf}
	\includegraphics[width=0.3\textwidth, page=13]{fig5_jack_coef_free.pdf}
	\includegraphics[width=0.3\textwidth, page=14]{fig5_jack_coef_free.pdf}
	\includegraphics[width=0.3\textwidth, page=15]{fig5_jack_coef_free.pdf}
	\includegraphics[width=0.3\textwidth, page=16]{fig5_jack_coef_free.pdf}
	\includegraphics[width=0.3\textwidth, page=17]{fig5_jack_coef_free.pdf}
	\includegraphics[width=0.3\textwidth, page=18]{fig5_jack_coef_free.pdf}
	\includegraphics[width=0.3\textwidth, page=19]{fig5_jack_coef_free.pdf}
	\includegraphics[width=0.3\textwidth, page=20]{fig5_jack_coef_free.pdf}
	\includegraphics[width=0.3\textwidth, page=21]{fig5_jack_coef_free.pdf}
	\caption{Empirical distributions of the Regional Ordinal Regression coefficients models, corresponding to minimum misclassification errors, obtained by a jackknife cross-validation procedure.}
	\label{fig:jack2}
\end{figure}

 
A reference model for each region, corresponding to the estimated model with the minimum misclassification error measured, has been selected and reported in the following tables. Table \ref{tab: regional models_info} collects the number of input variables, number of principal components, cumulative percentage of total variance explained by principal components, misclassification errors. The minimum misclassification errors correspond to models where the number of input variables varies from 1 to 8, the number of principal components varies from 1 to 4 without losing much information in terms of total variance explained ($> 92 \%$). \\
\begin{table}[!h]
	\scriptsize
	\caption{Regional model information corresponding to minimum misclassification errors}
	\label{tab: regional models_info}
	\centering
	\begin{tabular}{lcccc}  
		\toprule Region & Number of & Number of & Cumulative \% Variance & Misclassification \\   
		&Input  Variables & PC's &  explained  by PC's & Error  (\%)\\   
		\midrule
		\midrule
		Abruzzo &   5 & 3 & 98.60 & 22.92 \\ 
		Basilicata &   5 & 3 & 98.05 & 14.58 \\ 
		Calabria &   2 & 2 & 100.00 & 27.08 \\ 
		Campania &   2 & 2 & 100.00 & 29.17 \\ 
		Emilia-Romagna &   2 & 2 & 100.00 & 18.75 \\ 
		Friuli Venezia Giulia &   4 & 2 & 96.86 & 22.92 \\ 
		Lazio &   4 & 3 & 96.85 & 29.17 \\ 
		Liguria &   3 & 3 & 100.00 & 16.67 \\ 
		Lombardia &   2 & 2 & 100.00 & 18.75 \\ 
		Marche &   1 & 1 & 100.00 & 27.08 \\ 
		Molise &   2 & 2 & 100.00 & 22.92 \\ 
		P.A. Bolzano &   8 & 3 & 96.43 & 18.75 \\ 
		P.A. Trento &   2 & 2 & 100.00 & 22.92 \\ 
		Piemonte &   2 & 2 & 100.00 & 29.17 \\ 
		Puglia &   4 & 2 & 92.89 & 27.08 \\ 
		Sardegna &   6 & 4 & 97.72 & 12.50 \\ 
		Sicilia &   3 & 2 & 96.25 & 16.67 \\ 
		Toscana &   3 & 2 & 92.26 & 12.50 \\ 
		Umbria &   1 & 1 & 100.00 & 18.75 \\ 
		Valle d'Aosta &   5 & 3 & 96.83 & 29.17 \\ 
		Veneto &   4 & 3 & 95.48 & 18.75 \\ 
		\bottomrule
	\end{tabular}
\end{table}
The input variables selected and combined in terms of principal component are displayed in tables \ref{tab: regional models_pca} and \ref{tab: regional models_pca2} with the corresponding loadings. In order to emphasize the importance of each variable in the prediction of risk level, figure \ref{fig:vars} shows the percentage of model inclusion of each variable among the best selected models. 
\begin{table}[!h]
	\scriptsize
	\caption{Regional Principal Component loadings corresponding to minimum misclassification errors (Continued).}
	\label{tab: regional models_pca}
	\centering
	\begin{tabular}{llrrrr}  
		\toprule 
		Region & Input & \multicolumn{4}{c}{Variable Loadings} \\   
		& Variables & $PC_1$ & $PC_2$ & $PC_3$ & $PC_4$ \\   
		\midrule
		\midrule
		Abruzzo&Hospitalized with symptoms & 0.53 & 0.09 & -0.35 &\\ 
		& ICU patients & 0.53 & 0.06 & -0.40 &\\ 
		&Cases confirmed by PCR & -0.38 & -0.47 & -0.77 &\\ 
		&Increase in total cases & 0.52 & -0.25 & 0.10 &\\ 
		&Increase in total swabs & 0.15 & -0.84 & 0.34 &\\ 
		\midrule
		Basilicata&Home quarantine & -0.48 & 0.17 & 0.48 &\\ 
		&  Confirmed cases & -0.49 & 0.16 & 0.48 &\\ 
		& Deaths & 0.35 & -0.72 & 0.60 &\\ 
		&Increase in total cases & -0.47 & -0.41 & -0.20 &\\ 
		&Increase in total swabs & -0.44 & -0.51 & -0.37 &\\ 
		\midrule
		Calabria&ICU daily admissions & 0.71 & 0.71 &&\\ 
		&  Home quarantine & 0.71 & -0.71 &&\\ 
		\midrule
		Campania&Cases confirmed by RAT & 0.71 & 0.71 &&\\ 
		&  Total swabs & 0.71 & -0.71 &&\\ 
		\midrule
		Emilia-Romagna&ICU patients & 0.71 & -0.71&& \\ 
		&  Increase in total cases & 0.71 & 0.71 &&\\ 
		\midrule
		Friuli Venezia Giulia&ICU daily admissions & 0.51 & -0.07 &&\\ 
		&  Home quarantine & 0.52 & -0.35 &&\\ 
		& Confirmed cases & 0.52 & -0.35 &&\\ 
		&Increase in total cases & 0.46 & 0.87 &&\\ 
		\midrule
		Lazio&ICU patients & 0.60 & -0.21 & 0.04 &\\ 
		&  Increase in total cases & 0.57 & 0.26 & -0.71 &\\ 
		& RAT & -0.43 & 0.64 & -0.36 &\\ 
		&Increase in total swabs & 0.36 & 0.69 & 0.61 &\\ 
		\midrule
		Liguria&ICU patients & 0.64 & -0.35 & 0.69 &\\ 
		&  Cases confirmed by RAT & 0.41 & 0.91 & 0.08 &\\ 
		& Increase in total cases & 0.65 & -0.23 & -0.72 &\\ 
		\midrule
		Lombardia&ICU patients & 0.71 & -0.71 &&\\ 
		&  Increase in total cases & 0.71 & 0.71 &&\\ 
		\midrule
		Marche & Increase in total cases & 1.00 &&&\\
		\midrule
		Molise&Increase in total cases & -0.71 & 0.71 &&\\ 
		&  Increase in total swabs & -0.71 & -0.71 &&\\ 
		\bottomrule
	\end{tabular}
\end{table}
\begin{table}[!h]
	\scriptsize
	\caption{(Continued) Regional Principal Component loadings corresponding to minimum misclassification errors.}
	\label{tab: regional models_pca2}
	\centering
	\begin{tabular}{llrrrr}  
		\toprule 
		Region & Input & \multicolumn{4}{c}{Variable Loadings} \\   
		& Variables & $PC_1$ & $PC_2$ & $PC_3$ & $PC_4$ \\   
		\midrule
		\midrule
		P.A. Bolzano&Hospitalized with symptoms & -0.40 & 0.26 & -0.04 &\\ 
		&ICU patients & -0.38 & 0.27 & 0.29 &\\ 
		&Cases confirmed by PCR & 0.38 & 0.32 & -0.06 &\\ 
		&Cases confirmed by RAT & 0.36 & 0.25 & 0.48 &\\ 
		&Increase in total cases & -0.29 & 0.39 & -0.65 &\\ 
		&People Tested & 0.41 & 0.21 & -0.21 &\\ 
		&Total swabs & 0.37 & 0.31 & -0.30 &\\ 
		&Increase in total swabs & -0.15 & 0.63 & 0.35 &\\ 
		\midrule
		P.A. Trento&ICU patients & -0.71 & -0.71 &&\\ 
		&  Increase in total cases & -0.71 & 0.71 &&\\ 
		\midrule
		Piemonte&Deaths & 0.71 & -0.71 &&\\ 
		&  RAT & 0.71 & 0.71 &&\\ 
		\midrule
		Puglia&Discharged healed & 0.54 & -0.05 &&\\ 
		&Increase in total cases & -0.44 & 0.82 &&\\ 
		&People Tested & 0.54 & 0.28 &&\\ 
		&RAT & 0.48 & 0.49 &&\\ 
		\midrule
		Sardegna&Hospitalized with symptoms & -0.49 & 0.21 & -0.30 & 0.37 \\ 
		&  ICU patients & -0.47 & 0.34 & -0.22 & 0.30 \\ 
		& Discharged healed & 0.51 & 0.19 & 0.10 & 0.41 \\ 
		&Cases confirmed by RAT & 0.45 & 0.38 & -0.10 & 0.46 \\ 
		&Increase in total cases & -0.22 & 0.54 & 0.80 & -0.12 \\ 
		&Increase in total swabs & 0.19 & 0.61 & -0.45 & -0.61 \\ 
		\midrule
		Sicilia&ICU daily admissions & -0.60 & 0.08 &&\\ 
		&  Home quarantine & -0.57 & 0.66 &&\\ 
		& Increase in total cases & -0.56 & -0.75 &&\\ 
		\midrule
		Toscana&Cases confirmed by RAT & 0.60 & 0.56 &&\\ 
		& Increase in total cases & -0.77 & 0.19 &&\\ 
		& Increase in total swabs & -0.22 & 0.81 &&\\ 
		\midrule
		Umbria& Increase in total cases & 1.00 &&&\\
		\midrule
		Valle d'Aosta&Hospitalized with symptoms & -0.54 & 0.22 & -0.31 &\\ 
		& ICU patients & -0.51 & 0.28 & -0.43 &\\ 
		& Deaths & 0.33 & 0.62 & -0.32 &\\ 
		& Increase in total cases & -0.42 & 0.40 & 0.78 &\\ 
		& PCR & 0.40 & 0.57 & 0.13 &\\ 
		\midrule
		Veneto&ICU patients & -0.41 & 0.59 & -0.67 &\\ 
		&Increase in total cases & 0.15 & 0.72 & 0.62 &\\ 
		&PCR & 0.67 & -0.12 & -0.32 &\\ 
		&Increase in total swabs & 0.60 & 0.35 & -0.26& \\ 
		\bottomrule
	\end{tabular}
\end{table}
\begin{table}[!h]
	\scriptsize
	\caption{Regional Ordinal Regression coefficients models corresponding to minimum misclassification errors.}
	\label{tab: regional models_or}
	\centering
	\begin{tabular}{lrrrrrr}  
		\toprule 
		Region & $\eta_1$ & $\eta_2$ & $PC_1$ & $PC_2$ & $PC_3$ & $PC_4$ \\   
		\midrule
		\midrule
		Abruzzo& -1.45& 6.39 &0.86&-1.90&1.99&\\
		Basilicata& 0.36& 18.95 &0.08&-1.48&-2.72&\\
		Calabria& -0.34& 4.19 &0.97&3.74&&\\
		Campania& -0.33& 2.83 &-0.09&-0.88&&\\
		Emilia-Romagna& -0.71& 5.29 &1.77&3.31&&\\
		Friuli Venezia Giulia& -1.59& 5.12 &1.09&3.13&&\\
		Lazio& -0.08& 3.53 &0.64&0.33&-1.97&\\
		Liguria& -0.38& 6.43 &1.99&1.57&-2.65&\\
		Lombardia& -1.44& 6.4 &2.02&4.15&&\\
		Marche& -1.26& 4.63 &2.43&&&\\
		Molise& -0.68& 20.15 &-0.84&3.86&&\\
		P.A. Bolzano& -1.17& 9.65 &-1.29&-0.39&-2.94&\\
		P.A. Trento& -1.93& 17 $e^5$ &-2.33&4.54&&\\
		Piemonte& -0.54& 2.61 &-0.08&0.95&&\\
		Puglia& -0.23& 2.88 &-0.62&1.71&&\\
		Sardegna& 0.35& 4.16 &0.03&0.19&1.40&-2.18\\
		Sicilia& 0.17& 5.05 &0.10&-3.14&&\\
		Toscana& -0.77& 4.52 &-2.20&0.42&&\\
		Umbria& -0.27& 3.92 &2.55&&&\\
		Valle d'Aosta& -0.14& 4.62 &-0.23&0.17&1.81&\\
		Veneto& 0.2& 5.11 &0.22&1.09&5.43&\\
		\bottomrule
	\end{tabular}
\end{table}
Table \ref{tab: regional models_or} collects the parameters of the corresponding ordinal regression models. To assess the stability and robustness of the proposed models, a rigorous jackknife cross-validation procedure is implemented. This approach involved repeatedly resampling the data 1000 times, where, for each iteration, one day was randomly removed from each week across all regions simultaneously. This resampling strategy aims to minimize potential temporal fluctuations and assess the model's sensitivity to slight changes in the data. Notably, the principal component transformation is held fixed throughout the resampling process to maintain consistency in the dimensionality reduction process and ensure comparability across iterations.
The jackknife cross-validation procedure is particularly valuable in evaluating the reliability of the proposed models, as it provides a systematic method for quantifying uncertainty in the estimated parameters. By generating empirical distributions for each estimated coefficient, this technique allows for the definition of robust empirical confidence intervals for each $\beta_i$. As displayed in Figure \ref{fig:jack2}, the limited variability observed in these empirical distributions highlights the stability of the model parameters, even when subjected to perturbations in the data.
Such an approach is essential for validating the proposed methodology, as it ensures that the model's predictive performance is not overly sensitive to minor changes in the data collection process. This validation process further supports the practical applicability of the model for public health decision-making, providing policymakers with reliable and interpretable estimates for regional colour assignments.

\section{Conclusions and future work}
\label{conclusion}
This study examines the data provided by the Italian Ministry of Health during the COVID-19 pandemic, focusing on the colour-coded classification system used to apply regional mobility restrictions. Preliminary analysis highlights the redundancy present in the Ministry's comprehensive models, suggesting that increasing the number of input factors can degrade the quality of information rather than improve it.\\
Assuming the validity of the official colour assignments, a refined approach is proposed that simplifies the framework by selecting only a limited set of relevant factors. This reduction aims to improve clarity in analyzing relationships between indicators and improve the robustness of the decision-making process. The proposed methodology offers regional models that determine colour classifications, and thus corresponding restrictive measures, based on a streamlined set of predictors. 
Moving forward, estimating an ordinal regression model using weekly regional data is suggested to handle model interpretability and reliability. However, this approach requires rigorous evaluation of model fit through techniques such as data partitioning into training and validation sets. Such an assessment would enable a systematic approach to balancing model simplicity and complexity, resulting in a well-fitted model that effectively supports policymakers in assigning appropriate classifications to individual regions.\\
The assumption of correctness in the colour assignments is acknowledged to be potentially unrealistic, particularly considering the challenges encountered during data collection. This assumption also presumes that the distribution of input factors remains stationary, implying that past patterns are reliable predictors of future conditions. \\
Future research could address these limitations by employing more advanced and flexible machine learning algorithms for hidden patterns detection, directly managing multicollinearity, and refining the variable selection process. By optimizing the model's predictive accuracy and robustness, such efforts could provide more reliable tools for public health decision-making.
\bibliography{paper-refs}

\end{document}